\documentclass[12pt]{article}
     \newlength{\dinwidth}                       
     \newlength{\dinmargin}                      
\def\Journal#1#2#3#4{{#1} {\bf #2}, #3 #4}

\def\NPB{{\em Nucl. Phys.} B}
\def\CPC{{\em Comp. Phys. Comm.}}
\def\PLB{{\em Phys. Lett.}  B}

\def\PRD{{\em Phys. Rev.} D}
\def\ZPC{{\em Z. Phys.} C}
\usepackage {graphics}          
\usepackage {epsfig}

\begin{document}
\vspace*{-1.8cm}
\begin{flushright}
\flushright{\bf LAL 00-03}\\
\vspace*{-0.3cm}
\flushright{January 2000}
\end{flushright}

\vspace*{5mm}
\begin{center}  
{\bf\large THE EXCESS OF POSITIVE CHARGES AND \\
\vspace*{1mm}
 THE LEADING PARTICLES IN DIS AT HERA}\\

  \vspace*{5mm}
 
{\bf Beno\^\i t DELCOURT\\ }
 \end{center}
\begin{center}
{\bf Laboratoire de l'Acc\'el\'erateur Lin\'eaire}\\
{IN2P3-CNRS et Universit\'e de Paris-Sud, BP 34, F-91898 Orsay Cedex\\
 Email: delcourt@lal.in2p3.fr.}\\
\end{center}

\vspace*{3mm}
\begin{abstract}
 The charge of the leading particle in DIS interactions is studied with the QCD models LEPTO, ARIADNE, RAPGAP and HERWIG. It is predicted to be preferentially positive, as is expected from the modified quark parton model and is studied over a wide range of $x$ and $Q^2$. 
\end{abstract}

\section{\Large \textbf{Introduction}}
In deep-inelastic lepton-proton scattering processes, a small excess of positively charged leading particles is expected owing to the valence quark structure of the proton. An interaction involving a valence $u$ quark is 8 times more likely than that involving a valence $d$ quark. No excess is expected from sea quark or gluon initiated interactions.\\ 
Since leading hadrons ``remember`` the charge of the struck quark, this gives  the experimentalist a probe to investigate the partonic structure of the proton. At HERA this method could be used to perform such studies over several orders of magnitude in $x$ and $Q^2$ and will complement existing studies based on jet measurements and  other hadronic final state variables[1][2]. 

\subsection{\Large \textbf{Definitions}}

To define an event charge, let us first determine a reference direction on which to project momenta. If the DIS scattering were simply an elastic deflection of a quark by a positron, this direction would simply be the final quark direction. But we know that the reaction is more complicated, involving gluons or even more than one quark (boson-gluon-fusion BGF). Nevertheless, from the deflected positron parameters (which, for simulated events are known without error) we define the 'hadronic axis' by assuming the scattering to be elastic, and using the deflected positron information.\\
 Alternative methods could be to use event shape information, such as the thrust axis, to define the hadronic system. This is beyond the scope of this paper and is not considered here.\\
Event charge studies have already been made in deep inelastic scattering [3] and in e+e- interactions at LEP[4].\\
 For this paper, the following definition of event charge is used: $Q={\sum_iP^{||}_i q_i}/{\sum_iP^{||}_i}$, where $q_i$ is the particle charge (-1 or +1) and $P^{||}_i$ is the projected momentum on the hadronic axis .The sums run over $i$ all particles which have positive $P^{||}_i$'s; Q is between -1 and +1. This was used by Aleph in [4].\\
        
\subsection{Relation of event charge Q to scattering on valence quarks.\\
The Leading particle and the positive excess $\varepsilon_+$ }
For this study, the leading particle is defined as the charged hadron which has the largest momentum projected onto the hadronic axis. If the experimentally measurable $K^0_s$ or $\Lambda$ were to be the leading particle then the event is discarded.

The positive excess $\varepsilon_+$ is defined as the ratio between the number of leading particles which are positively signed $N_+$ over those which are negatively signed $N_-$: $\varepsilon_+=\frac{N_+}{N_-}$.
\subsubsection{Relation of $\varepsilon_+$ to interactions of valence quarks}
In this section, we will use $\varepsilon_+$ to explain the relation of the positive mean event charge to interactions involving the proton valence quarks.
\vspace*{-1mm}
The magnitude of  $\varepsilon_+$ can be related to two quantities:
\vspace*{-2mm}
\begin{itemize}
\item[$\triangleright$ $\varepsilon^+_q$:] the positive excess at the quark level, given by the ratio $\varepsilon^+_q=\frac{M^+}{M^-}$ where $M^-=N_{\overline{u}}+N_d+N_s$ and $M^+=N_u+N_{\overline{d}}+N_{\overline{s}}$;   $N_u$, $N_d$, $N_s$, $N_{\overline{u}}$, $N_{\overline{d}}$
 and $N_{\overline{s}}$ are the numbers of those events where the interaction has given a leading u-,d-,s-,$\overline{u}$-,$\overline{d}$-, respectively $\overline{s}$-quark.
\vspace*{-2mm}
\item [$\triangleright$ p:] the probability that the charge of the initial quark and  the one of the leading particle are the same. Here $p$ should be the same irrespective of the quark flavour. This will later be shown a flawed assumption, but we can take mean values.
\end{itemize}
\vspace*{-2mm}
\noindent
Then the dependence of $\varepsilon^+$ upon $\varepsilon^+_q$ and $p$ is described by the formula:
\vspace*{-1mm}
$$\varepsilon^+={(pM^+(1-p)M^-)}/{(pM^-+(1-p)M^+)}={(1-p+\varepsilon^+_qp)}/
{(p+\varepsilon^+_q(1-p))}. $$
\vspace*{-7mm}

The probability $p$, averaged on different flavours, is given by QCD based models, and is found to be of the order of 0.6 to 0.7.\\
 The valence quark interactions from the proton are 8 times more likely to be initiated by an $u$ quark than by a $d$ quark . Hence, for valence quarks, more $\pi^+$ and $K^+$ than $\pi^-$ and $K^-$ should be detected as leading particles.\\
 For example, in the very high $x$ region, the valence quark 
interactions will dominate the $ep$ cross-section. Therefore, $\varepsilon^+_q$ is equal to 8, and $\varepsilon_+$ is expected to be 1.39 for $p=0.6$.\\
    To quantify this, we take $Y^+_{\alpha}$ and  $Y^-_{\alpha}$ as the yields of positive and negative quarks struck by the photon, respectively, and produced by some process $\alpha$. To simplify, we could  consider only three different processes $\alpha$: Boson Gluon fusion (BGF), scattering on a sea quark (Sea) or scattering on valence quarks (V).\\
 Here, we do not distinguish between elastic $e-q$ diffusion and QCD Compton reactions, where one gluon is added: $e^++q\longrightarrow e'^+ +q'+gluon$. It has been shown with LEPTO, that so far as the event charge is concerned, they are equivalent. This can be simply understood: in QCD compton events, the gluon is often weaker than the final quark q', and also it has to break down into two quarks (plus gluons). The chance that one out of these two quarks has a larger energy than q' is very limited; so the leading particle will generally stay related to the q' quark.\\
The positive excess at the quark level is thus: 
 $\varepsilon^+_q={(Y^+_{BGF}+Y^+_{sea}+Y^+_V)}/{(Y^-_{BGF}+Y^-_{sea}+Y^-_V)}$.\\
For sea quarks and quarks emitted through BGF, the number of quarks is identical to the number of their antiquarks, we can state:\\
\vspace*{-2mm}
        $$Y^+_{BGF}+Y^+_{sea}=Y^-_{BGF}+Y^-_{sea}=Y_0.$$\\Then: $\varepsilon^+_q={(1+\frac{Y^+}{Y_0})}/{1+\frac{Y^-_V}{Y_0})}$.
Let $R=\frac{Y^+_V}{Y^-_V}$ and $S=\frac{Y^+_V}{Y_0}$, so that $\frac{Y^-_V}{Y_0}=\frac{S}{R}$. As was quoted above, the value of R should be 8 (naively), so that :  $\varepsilon^+_q={(1+S)}/{(1+\frac{S}{R})}$ is nearly equal to $1+S$.
We see then that the positive excess at the quark level is a measure of the ratio of yield by valence quarks over yield by other processes.\\
 We access to it by measuring the mean event charge, but the probability $ p$ introduced above should be known, using QCD based models.  We can perform some tests on these QCD based models, mainly on various fragmentation properties [5] and leading particle studies (mainly $K_0^S$ and $\Lambda$ [6]). After determining the mean event charge as a function of x and $Q^2$, we cannot have direct values of $\varepsilon^+_q $ but we may compare the reconstructed charges with those predicted by QCD based models. 
\vspace*{-2mm}
\subsection {Event selection}
\vspace*{-3mm}
The leading  particle plays a major role in determination of the mean charge, and so we will perform our kinematic cuts on the leading  particle parameters. In the following, we will cut all events having a leading particle angle less than 0.7 rad (where the angle is defined with respect to the initial proton direction ) or larger than $\pi-0.7$ rad. In the forward direction, we then avoid particles coming from the proton remnant. In both directions, we have to reserve some room for the next to leading (n.t.l.) and other particles, so that the event charge is well evaluated.\\
Another cut on all tracks has been imposed by experimental conditions at H1: to avoid beam gas background, to which the mean value of the event Charge Q is very sensitive, we require all tracks to have a transverse momentum larger than 0.6 GeV/c. \\
The parameter  $ f={P^{||}_{leading}}/{P_{quark}}$ was also used to select events where the leading particle ``remembers`` better the initial quark flavour. Typically, it was required to be larger than 0.15.
\vspace*{-8mm}
\subsection {Our sample of events}
\vspace*{-3mm}
        There exists different QCD based models, for which we take the following versions: LEPTO(6.5)[7], ARIADNE(410)[8], RAPGAP(2.0)[9], and HERWIG(5.9008)[10]. LEPTO, ARIADNE, and RAPGAP have the same hadronisation program, JETSET [11].\\
For each of these four QCD based models, we have generated 2 million events in the range $3.<Q^2<100$ GeV$^2$. The corresponding luminosity is $6.3$ pb$^{-1}$, which is below what is now available at HERA. The GRV [12] parton density was used for all. For ARIADNE and RAPGAP, the pomeron was off. For LEPTO, the QCD effects were on.
\vspace*{-4mm}
\section{Fragmentation studies}
\vspace*{-2mm}
 We have made  comparisons of different QCD based models for the following fragmentation parameters:
\vspace*{-3mm}
\begin{itemize}
\item[-] charge correlations in the system of the 2 or 3 leading particles;
\vspace*{-3mm}
\item[-] projected momentum of the leading particle on the hadronic direction;
\vspace*{-3mm}
\item[-]$\varphi$, rapidity and momentum ratio correlations of the leading and the struck quark;
\vspace*{-3mm}
\item[-]$\varphi$, rapidity and charge correlations of the leading and n.t.l. particles;
\vspace*{-3mm}
\item[-]$K^0_S$, $\Lambda$ as leading particles;
\vspace*{-3mm}
\item[-]$\rho$, $K^{\star\pm}$ and $\phi$ production in the system of the leading and next to leading particles.
\end{itemize}

 We show here charge correlations in the system of the 2 or 3 leading particles. However, the other parameters are shown in the Internet version of this note. \\ 
All these quantities can be experimentally measured, and comparison to M.C's provides a good test of hadronisation programs.The detailed results are shown in the extended version (on the web). Generally, there is a good agreement between different QCD based models.
\subsection{ Charge correlations for leading and n.t.l. particles}
The leading and next to leading particles may either have opposite charges, which is foreseen by the naive model of quark hadronisation, or same charges, which is also foreseen but in less frequent cases. In the following, when we write a pair of signs, the first is for the leading, the second for the n.t.l.. In the next figure 4 different ratios are shown, for channels 1 to 4:

\begin{itemize}
\item [1:] $\frac{+- and - -}{  ++ and -+}$: this is the 'negative excess' of the n.t.l. Naively, this should be a little larger than 1, and it is very close to 1 for all QCD based models.
\item [2:] $\frac{++}{ - -}$: this is the positive excess for leading and n.t.l. when both have the same charge.
\item [3:] $\frac{+-}{ - +}$: positive excess for opposite charge leading and n.t.l, lower than the preceding one.
\item [4:] $\frac{+ + +}{ - - -}$: this is the positive excess when the three first particles carry the same sign.
\end{itemize}

\begin{figure}[h]
\centering
\vspace*{-8mm}
\includegraphics*[width=15cm,height=12cm,clip=true]{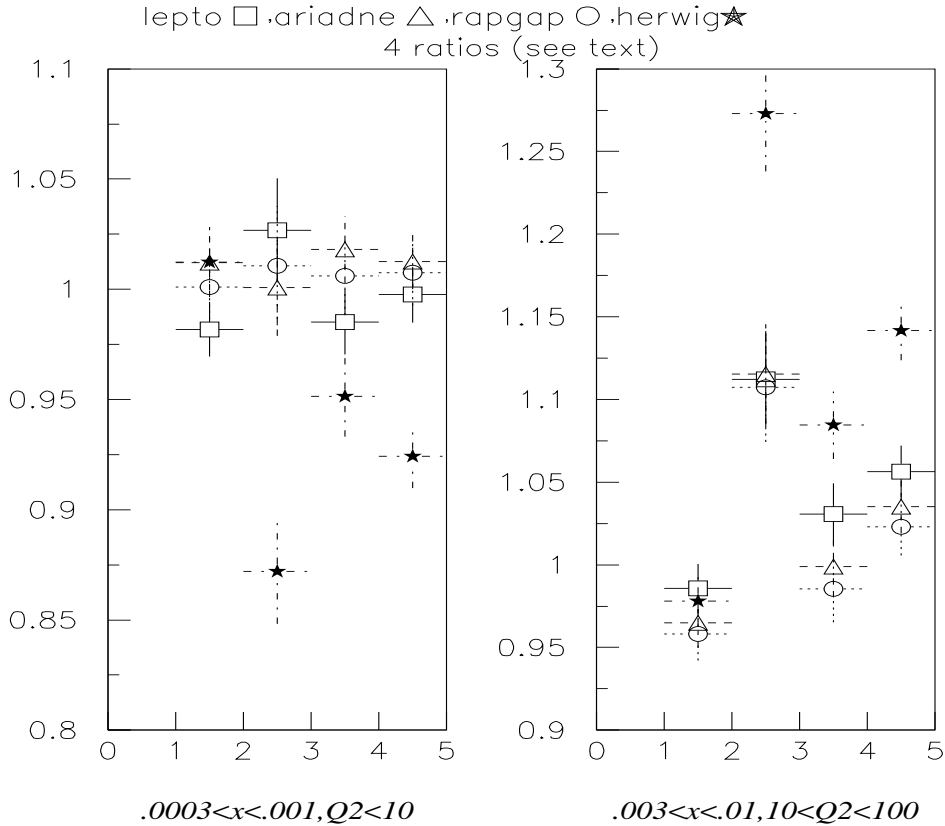}
\vspace*{-2mm}
\caption{4 ratios (defined in the text) for various x and $Q^2$ and for $ f=\frac{P^{||}_{leading}}{P_{quark}}>0.15$.}
\end{figure}

The agreement between the three JETSET  QCD based models and on the other hand HERWIG is rather poor.\\ 

In the next figure, we show 3 other ratios:
\begin{itemize}
\item[1:] $\frac{\rm different\ signs }{\rm same \ signs}$ for the leading and the n.t.l..It should be much larger than 1, and it is of the order of 2. 
\item[2:] $\frac{\rm different \ signs }{\rm same \ signs}$ for the leading and the next to next to leading (n.t.n.t.l.). Naively, this would be expected to be less than 1 but is larger than 1. This is due to the fact that events with signs (+++) or (- - -) are very seldom, as is shown in the next channel.
\item[3:] $\frac{(++ -) + (+ - +)+ (+ - -) +(- -  +) +(- +  -)+(- + +)}{3 *( (+++)+(- - -) )}$.\\We see that the value of this ratio is between 3 to 4: it is very seldom that the three first particles carry the same sign. 
\end{itemize}
\vspace*{5mm}
\begin{figure}[ht]
\vspace*{5mm}
\centering
\includegraphics*[width=15cm,height=12cm,clip=true]{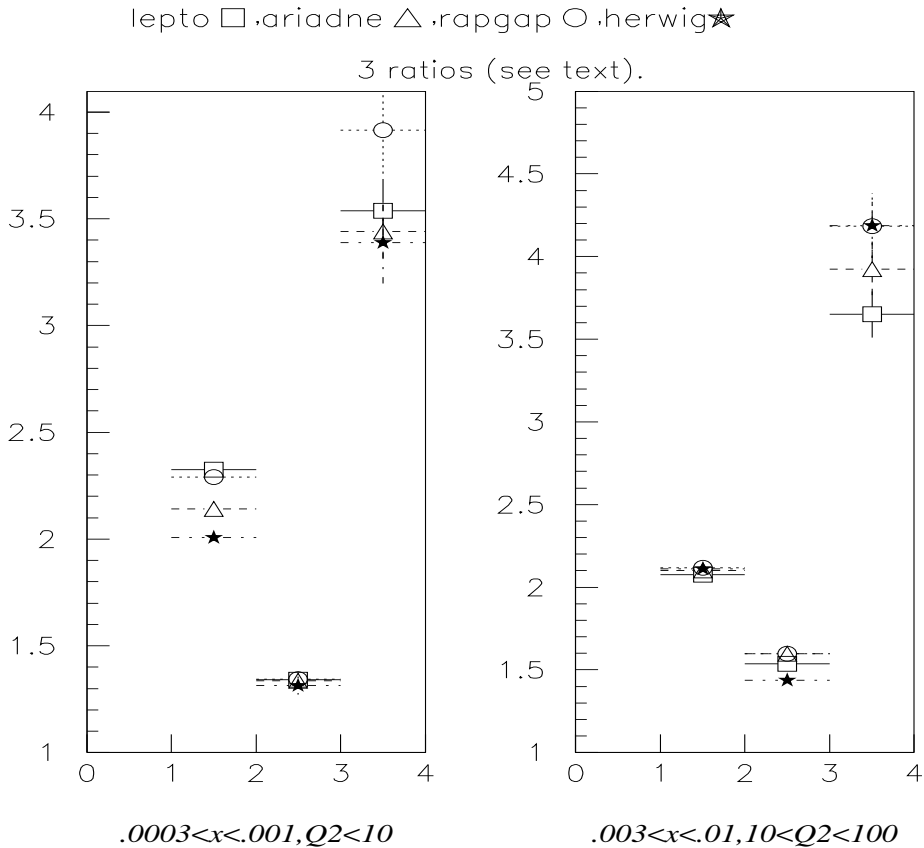}
\vspace*{8mm}
\caption{3 ratios (defined in the text just above)for some x and $Q^2$ and for $ f=\frac{P^{||}_{leading}}{P_{quark}}>0.15.$}
\end{figure}
\newpage
\section{QCD model predictions for different parton induced quantities}
As was written in the introduction, the event charge Q depends upon two quantities: the mean probability $p$ that the initial quark charge and  that of the leading particle are the same and $\varepsilon^+_q$: the positive excess at the quark level. Before presenting the results on the event charge, we will present the predictions of different QCD based models for these quantities.\\
\subsection{The sign excess depending on the flavour}
 The sign excess related to a quark is defined as the ratio  of the number of events originated from that quark, which gives a leading particle having its sign, to the number of those which originate from $q$ as well, but give a leading particle of opposite sign. For instance, for the u quark initiating the reaction, it is: $S_u=\frac{N_+}{N_-}$, whereas for the $\overline{u}$:$S_{\overline{u}}=\frac{N_-}{N_+}$. The sign excess is related directly to the probability
 $p$: for example, if we have a sign excess of 3, then the probability $p$ for that quark is 75 percent. Here we compare the $S_q{\rm '}s$  for the 4 QCD based models.

\begin{figure}[h]
\centering
\vspace*{-2mm}
\includegraphics*[width=15cm,height=11cm,clip=true]{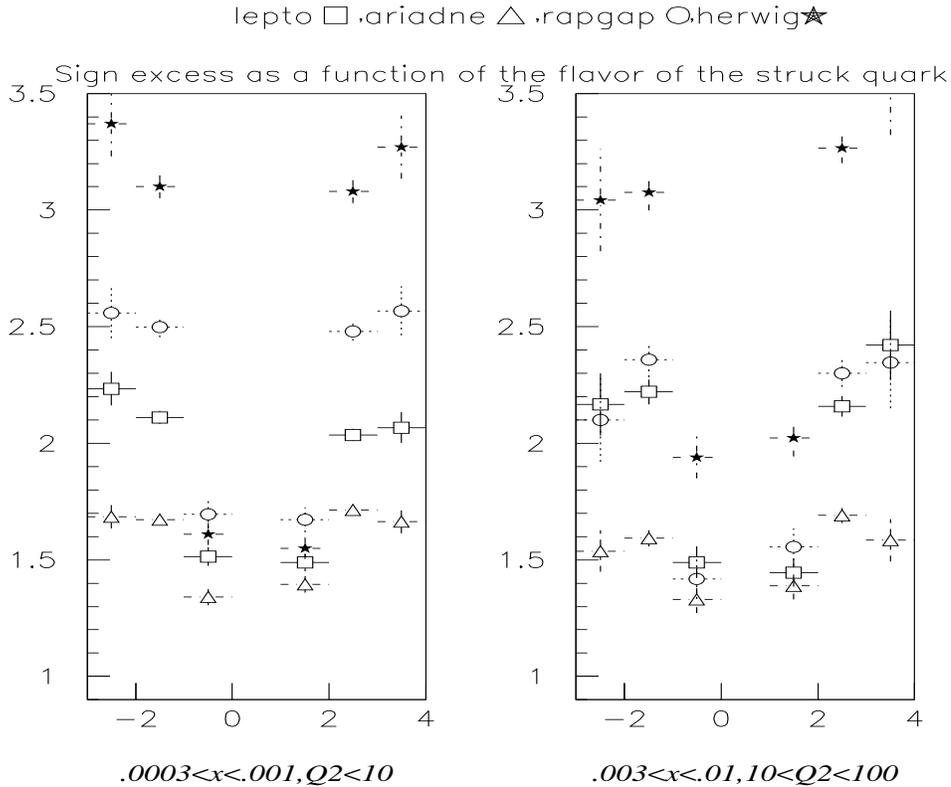}
\caption{ Sign excess as a function of the flavour of the struck quark, for some $x$ and $Q^2$.}
\end{figure}
Each box of the preceding figure has 6 channels, from -3 to 3. We use the standard quark identity of the Particle Data Group [13]:-3 is $\overline{s}$, -2 is $\overline{u}$, -1 is $\overline{d}$, 1 is $d$, 2 is $u$ and 3 is $s$. The sign excesses are shown for these 6 abscissa of the boxes.\\
 It is easily seen that the $d$,$\overline{d}$ have less sign excess than $u$,$\overline{u}$; there are at least 2 reasons: first, in the naive model, the
 $d$ can give a proton as leading particle,which then has the wrong sign; second, for what concerns charged particles, the $u$ can give a $\pi^+$ or a $K^+$, the $d$ only a  $\pi^-$ and not a $K^-$ [14], and the $u$ has there also an advantage.\\
It is seen that the sign excesses for a quark and its antiparticle are equal (within statistical errors), which is a good consistency check of the models.\\
Generally, HERWIG gives a greater sign excess than RAPGAP and LEPTO, which in turn gives more than ARIADNE. Let us point out that the sign excess reflects on one hand the hadronisation, (production of mesons and baryons), and on the other hand the production of gluons and quarks from an initial quark (the one from the proton) . As ARIADNE uses the same hadronisation program as RAPGAP and LEPTO, we see that there is a sensitivity to the underlying differences in the physics of these models. 
\subsection{The positive excess at the level of initial quarks}
The positive excess at the level of quarks: $\varepsilon^+_q={(N_u+N_{\overline{d}}+N_{\overline{s}})}/{(N_{\overline{u}}+N_d+N_s)}$, where $N_q$ is the number of events where  a quark $q$ from the proton initiates the reaction. On the next plot is shown how this quantity varies with $x$ and $Q^2$.
We can see that, as expected, $\varepsilon^+_q$ increases with $x$ (more valence at higher $x$), and that, at large $Q^2$, it begins to take important values.\\
The different QCD based models show similar behaviours, except for HERWIG for\linebreak $Q^2<10$ GeV$^2$.
\begin{figure}[ht]
\centering
\includegraphics*[width=15cm,height=11cm,clip=true]{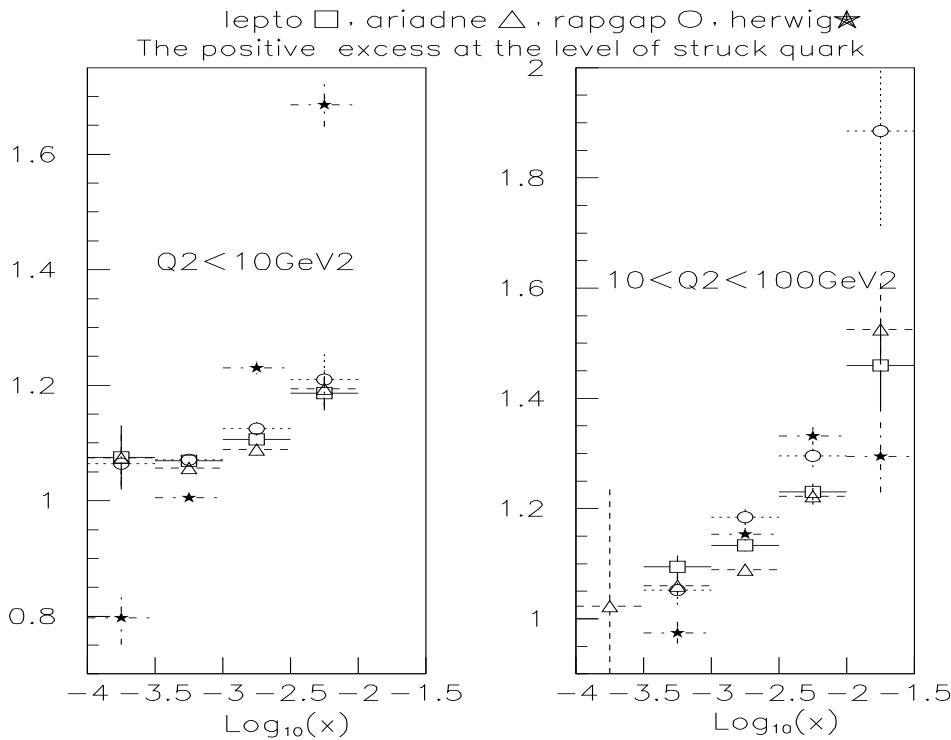}
\caption{ The positive excess at the level of initial quarks as a function of $log_{10}x$ and $Q^2$.}
\end{figure}
\subsection{Final result: mean event charge as a function of x and $Q^2$}
To recap, the charge that we consider here is the following: $Q=\frac{\sum_iP^{||}_i q_i}{\sum_iP^{||}_i}$, where the sum over $i$ extends to all particles having a positive momentum projection on the struck quark direction.\\
 The found mean event charges are shown in the next figure.
HERWIG predicts a far greater charge than the other models.
The charge increases with $x$ for the two ranges in $Q^2$ considered. 

\begin{figure}[ht]
\centering
\includegraphics*[width=15cm,height=12cm,clip=true]{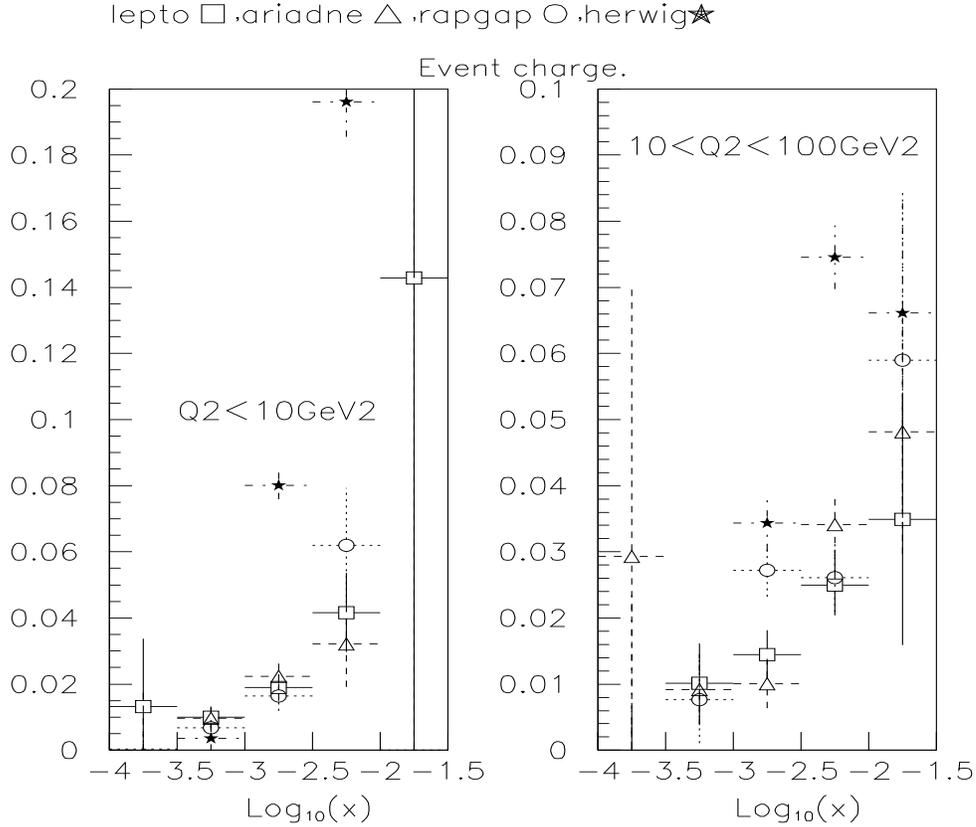}
\caption{Event charge  as a function of $log_{10}x$ and $Q^2$ for events within the cuts 
(mainly \break $0.7<$ leading particle $\theta <\pi-0.7$).}
\end{figure}
\section{Conclusions}



 Using four QCD based models the charge properties of inclusively
 produced hadrons in $ep$ DIS processes has been investigated.
Deviations of these models have been seen, particularly between those
 models employing string and cluster hadronisation for the total event
charge. Especially, HERWIG seems to give very different results than
 LEPTO, ARIADNE and RAPGAP, which all three use the hadronisation program
 JETSET.
\newpage

The mean event charge may also be measured in special cases of
 deep inelastic scattering:
\begin{itemize}
\item[-i-] for charge current events, where it should have the sign of the
 incoming electron or positron in the beam. Due to limited statistics,
this could
 only be an interesting check.\\
 \item[-ii-] for diffractive events. In the Pomeron model, the mean event
 charge is obviously zero. If we find a mean positive charge in these events,
 this means either that there is another mechanism involved in the
 diffraction or that the cuts selecting these diffractive events are not
 severe enough.
\end{itemize}
Michel Jaffr\'e and Ursula Berthon are acknowledged for their help in
processing data.


\begin{thebibliography}{99}
\bibitem{label1}
The H1 collaboration,  \Journal{\ZPC}{70} {(1996)} {609}.
\bibitem{label2}
The H1 collaboration, \Journal{\PLB}{428}{(1998)}{206}.
\bibitem{label3}
 J.P. Albanese et al. (the E.M.C. collaboration), \Journal{\PLB}{ 144}{(1984)}{302}. 
\bibitem{label4}
 The ALEPH collaboration, \Journal{\ZPC}{71}{(1996)}{357}.
\bibitem{label5}
The H1 collaboration, \Journal{\ZPC} {63}{(1994)}{377}.
\bibitem{label6}
The H1 collaboration, \Journal{\NPB} {480}{(1996)}{3}.
\bibitem{label7}
G. Ingelman, Proceedings of the workshop Physics at HERA, vol 3, Eds. W. Buchmueller and
 G. Ingelman, DESY (1992) 1366.
\bibitem{label8}
L. Loennblad, \Journal{\CPC} {87}{(1992)}{15}.
\bibitem{label9}
H. Jung, \Journal{\CPC} {86}{(1995)}{147}.
\bibitem{label10}
G. Marchesini et al., \Journal{\CPC} {67}{(1992)}{465}.
\bibitem{label11}
T. Sjoestrand \Journal{\CPC} {82}{(1994)}{74}.
\bibitem{label12}
M. Glueck, E. Reya, and A. Vogt, \Journal{\ZPC}{53}{(1992)}{127}.
\bibitem{label13}
Particle Data Group, see \Journal{\PRD} {54}{(1996)}{1}.
\bibitem{label14}
Ian Knowles, private communication.

\end{thebibliography}
\end{document}